\documentclass[twocolumn]{revtex4-2}

\usepackage{graphicx}
\usepackage{amsmath,amssymb}
\usepackage{color}
\usepackage{hyperref}
\usepackage{placeins}
\usepackage{float}
\usepackage{appendix}
\usepackage{multirow}
\usepackage{soul}
\usepackage{booktabs} 

\begin{document}
	
	\title{Thermo-coalescence model for Light Nuclei production in Relativistic Heavy-Ion Collisions}

	\author{Nachiketa Sarkar}
	\thanks{nachiketa.sarkar@gmail.com }
	\affiliation{Department of Physical Sciences, Indian Institute of Science Education and Research Berhampur, Laudigam - 760003, Ganjam, Odisha, India}
	\affiliation{Department of Physics, Amrita School of Physical Sciences, Coimbatore, Amrita Vishwa Vidyapeetham, India}

	\date{\today}
	\begin{abstract}
		We perform a Bayesian calibration of the Cross-term Excluded-Volume Hadron Resonance Gas (Cross EV--HRG) model, which incorporates flavor-dependent repulsive interactions within a thermodynamically consistent framework. For the first time, the thermal model is simultaneously constrained using lattice QCD (LQCD) thermodynamic observables and centrality-resolved hadron yield data from Pb–Pb collisions at $\sqrt{s_{\mathrm{NN}}}=2.76~\mathrm{TeV}$ measured by the ALICE Collaboration. We also find that the calibration outcome is strongly data-dependent in terms of constraining power and uncertainty structure. In particular, LQCD observables alone provide only weak constraints on the eigenvolume parameters, while the inclusion of hadron yield data substantially enhances the constraining power and induces a nontrivial reshaping of the posterior distributions. We further investigate the impact of correlated experimental systematic uncertainties by constructing a phenomenological covariance matrix and systematically varying its strength, demonstrating that a careful and consistent treatment of systematic correlations is essential for reliable parameter estimation. Across all calibration scenarios, the parameters associated with multi-strange hadrons remain only moderately constrained, which may reflect limitations of the currently established hadron resonance spectrum. No clear monotonic hierarchy of strange-hadron eigenvolume radii emerges within the present uncertainties, indicating that further dedicated studies are required.
	\end{abstract}

	\maketitle
\section{Introduction}
    
	The Hadron Resonance Gas (HRG) model, which treats hadrons and resonances as non-interacting thermal particles~\cite{Braun-Munzinger:1994ewq,Cleymans:1999st,Braun-Munzinger:2014lba,Alba:2017mqu}, has demonstrated its reliability by successfully reproducing LQCD thermodynamic observables below the pseudo-critical temperature~\cite{Borsanyi:2010cj,Borsanyi:2011sw,Borsanyi:2013bia,HotQCD:2014kol,Fodor:2018wul}. The HRG model is also extensively applied to extract chemical freeze-out parameters—namely the chemical freeze-out temperature $T_{\mathrm{ch}}$ and the baryon, strangeness, and electric-charge chemical potentials $\mu_{B}$, $\mu_{S}$, and $\mu_{Q}$—through comparisons with measured hadron yields, yield ratios, and fluctuation observables across a wide range of collision energies in heavy-ion experiments at the SIS, AGS, SPS, RHIC, and the LHC~\cite{Cleymans:1992zc,Cleymans:1998yb,Yen:1998pa,Cleymans:2005xv,Becattini:2005xt,Andronic:2005yp,Vovchenko:2015cbk,Andronic:2017pug,STAR:2017sal,Bellwied:2018tkc,Vovchenko:2018fmh,Poberezhnyuk:2019pxs,Bhattacharyya:2019cer,Bhattacharyya:2020sgn}. These freeze-out parameters are essential for determining the chemical composition of the fireball and for mapping the QCD phase diagram. The thermal statistical model, like the HRG model, is expected to play an equally important role at upcoming facilities such as NICA~\cite{Kekelidze:2016hhw} and FAIR~\cite{Friman:2011zz}.
	
    However, the ideal HRG model neglects short-range repulsive interactions, which become increasingly relevant at the particle densities characteristic of the hadronic phase near chemical freeze-out. Various formulations of the interacting HRG model have therefore been proposed in the literature to better capture the underlying physics~\cite{Yen:1997rv,Braun-Munzinger:1999hun, Begun:2012rf, Bugaev:2013sfa, Vovchenko:2014pka, Albright:2014gva, Satarov:2014voa, Alba:2016fku, Alba:2016hwx, Vovchenko:2016ebv, Vovchenko:2016rkn, Huovinen:2017ogf, Sarkar:2017bqy, Sarkar:2017ijd, Alba:2017bbr, Vovchenko:2017zpj, Dash:2018can, Sarkar:2018mbk, Motornenko:2020yme}.
    
    Numerous studies have shown that incorporating excluded-volume effects substantially improves both the description of LQCD thermodynamics and the quality of thermal fits to hadron yield data~\cite{Yen:1998pa, Andronic:2012ut,Vovchenko:2014pka,Vovchenko:2015cbk,Sarkar:2017ijd}. At the same time, the determination of chemical freeze-out parameters, as well as the agreement of thermodynamic observables with LQCD simulations, remains highly sensitive to the specific formulation of the HRG model and the parametrization of repulsive interactions~\cite{Vovchenko:2015cbk, Vovchenko:2016ebv}. Consequently, the treatment of eigenvolume interactions is a central, yet nontrivial, aspect of EV--HRG modeling. A commonly used approximation assumes a uniform hard-core radius for all hadron species~\cite{Yen:1997rv,Braun-Munzinger:1999hun,Vovchenko:2014pka,Sarkar:2017bqy,Sarkar:2017ijd}, which simplifies the thermodynamics but leads to a near-cancellation of EV effects when yield ratios are used for freeze-out parameter extraction. More sophisticated prescriptions assign species-dependent eigenvolumes by differentiating mesons from baryons~\cite{Begun:2012rf,Bugaev:2013sfa}, implementing mass-dependent radii~\cite{Albright:2014gva,Alba:2016fku}, or by parametrizing the eigenvolume in terms of the hadron’s internal structure—either through flavor content (light versus strange)~\cite{Alba:2016hwx,Alba:2017bbr} or the number of constituent quarks~\cite{Motornenko:2020yme}. In particular, Ref.~\cite{Alba:2016hwx} introduced a flavor-sensitive parametrization in which the eigenvolume of light (non-strange) hadrons grows proportionally with their mass, while that of strange hadrons decreases inversely with mass. This hybrid formulation significantly improved the quality of hadron yield fits across collision energies while remaining compatible with LQCD thermodynamic constraints. Recent and more general EV--HRG extensions include virial-expansion formulations with inter-species cross terms~\cite{Gorenstein:1999ce,Vovchenko:2016ebv,Satarov:2016peb,Vovchenko:2017zpj}, commonly referred to as the Cross EV--HRG scheme. A comprehensive analysis of the sensitivity of thermal fits to eigenvolume parametrization within this framework was presented in Ref.~\cite{Vovchenko:2016ebv}, which also explored a bag-like eigenvolume prescription calibrated to reproduce the proton’s ground-state radius. Beyond yield-based analyses, complementary constraints on effective hadron radii arise from comparisons with LQCD fluctuation observables. Recent studies~\cite{Karthein:2021cmb} indicate that the inclusion of additional high-mass resonances—beyond those listed in the standard PDG compilations—may be essential for accurately reproducing key features of the fluctuation data.
	Together, these studies demonstrate that the extracted freeze-out conditions and inferred eigenvolume parameters depend sensitively on both the chosen EV prescription and the dataset used for calibration~\cite{Vovchenko:2015cbk,Vovchenko:2016ebv,Alba:2016hwx,Alba:2017bbr}. In addition, methodological choices—such as using yields versus ratios, constructing ratios in different ways, or adopting particular treatments of systematic uncertainties—can also lead to significant variations in the inferred thermal parameters~\cite{Andronic:2005yp,Becattini:2007wt,STAR:2017sal,Bhattacharyya:2019cer,Bhattacharyya:2020sgn}. Nevertheless, most existing analyses still rely on $\chi^2$ minimization applied to only one of the two principal datasets—either LQCD thermodynamic observables or experimental data—leaving open the question of whether a single, self-consistent set of eigenvolume parameters can simultaneously describe both. Furthermore, earlier freeze-out studies typically fitted each centrality class independently~\cite{Becattini:2014hla,Cleymans:2004pp,STAR:2017sal,Alba:2016hwx}, thereby neglecting correlated systematic uncertainties shared across centralities and particle species. These limitations motivate the development of a unified, statistically robust framework that can consistently integrate theoretical and experimental constraints.
	
	The present work addresses these limitations by performing the simultaneous Bayesian calibration of a thermal model—the Cross EV--HRG framework—against both continuum-extrapolated LQCD thermodynamic observables and centrality-resolved ALICE hadron yields at $\sqrt{s_{NN}} = 2.76~\mathrm{TeV}$. The central novelty of this analysis lies in the unified and statistically rigorous treatment of these complementary datasets. This hybrid approach enables a direct assessment of whether a single set of model parameters can simultaneously describe the bulk thermodynamic properties extracted from LQCD within the hadronic phase up to the pseudo-critical temperature, and the experimentally measured particle composition at chemical freeze-out. Another key feature of our analysis is the explicit treatment of correlated systematic uncertainties through a phenomenologically motivated covariance matrix. To our knowledge, this is the first HRG study to incorporate such correlations in a fully Bayesian framework, enabling a quantitative assessment of their impact on parameter stability, degeneracies, and physical interpretation. To clarify the relative constraining power of the different inputs, we also perform independent calibrations using subsets of the data. Overall, the Bayesian framework yields reliable parameter estimates, more realistic constraints on hadronic eigenvolumes and freeze-out conditions, and a clearer characterization of parameter-space correlations and degeneracies within a single, coherent, and statistically robust analysis.

	The paper is organized as follows. In Section~\ref{sec:model_description}, we describe the Cross EV--HRG model. Section~\ref{sec:analysis_methodology} outlines the Bayesian analysis methodology. Results and discussion are presented in Section~\ref{results_discussion}, followed by summary and conclusions in Section~\ref{summary_conclusions}.
		
\section{Model Description: Cross EV--HRG Model}
	\label{sec:model_description}
	
	The ideal HRG model has been extended by incorporating excluded-volume (EV) corrections to account for the finite spatial extent of hadrons in a thermodynamically consistent manner~\cite{Rischke:1991ke,Yen:1997rv}. In this framework, each hadron species is assigned an eigenvolume, effectively modeling the short-range repulsive interactions.  
	
 	However, conventional EV--HRG formulations typically neglect the cross terms in the virial expansion that encode interactions between different hadron species. In contrast, the Cross EV--HRG model explicitly incorporates these inter-species virial contributions, in a thermodynamically consistent way. Consequently, it provides independent control over meson–meson, baryon–baryon, and meson–baryon interactions, and can also reflect the effective suppression of repulsive particle–antiparticle interactions~\cite{Satarov:2016peb} without introducing additional dynamical parameters. Although the inclusion of cross terms leads to coupled nonlinear equations, the Cross EV--HRG scheme respects the second-order virial expansion for a multi-component hadron gas, providing a physically grounded description of short-range repulsive interactions~\cite{Gorenstein:1999ce,Vovchenko:2016ebv,Vovchenko:2017zpj,Satarov:2016peb}.
	\vspace{0.2cm}
	\noindent The partial pressure in the Cross EV--HRG model is given implicitly by~\cite{Gorenstein:1999ce,Vovchenko:2016ebv}:
	\begin{equation}
		P_i(T,\mu_i) = P^{\text{id}}_i\left(T, \mu_i - \sum_{j \in H} \tilde{b}_{ij} P_j \right).
	\end{equation}
	Here, \(T\) denotes the temperature, \( \mu_i \) is the chemical potential of species \(i\), and \(H\) denotes the full set of hadronic species. The term \( \tilde{b}_{ij} \) denotes the effective second-order virial coefficient associated with hard-sphere repulsion and is defined as:
	\begin{equation}
		\tilde{b}_{ij} = \frac{2b_{ii}b_{ij}}{b_{ii}+b_{jj}}. 
	\end{equation}
	The coefficient \( b_{ij} \) represents the geometric excluded-volume characterizing interactions between species \( i \) and \( j \). It is defined as:
	\begin{equation}
		{b}_{ij}    = \frac{2 \pi}{3} (r_i+r_j)^3 ,
	\end{equation}
	where \( r_i \) denotes the hard-core radius of hadron species \( i \). The ideal-gas pressure $P^{\text{id}}_i\left(T, \mu^*_i\right) $  is evaluated at the effective chemical potential, $\mu^*_i$, which incorporates the repulsive interaction effects and is defined as:
	\begin{equation}
		\mu_i^* = \mu_i - \sum_j  \tilde{b}_{ij} P_j.
	\end{equation}
	The full system of coupled nonlinear equations is solved iteratively using Broyden’s method.
	
	\vspace{0.2cm}
	\noindent
	Thermodynamic consistency leads to the following relation for the number density of species $i$~\cite{Vovchenko:2016ebv}:
	\begin{equation}
		T n_i + P_i \sum_{j \in H} \tilde{b}_{ij} n_j = P_i.
	\end{equation}
	Once the pressure is determined self-consistently, other thermodynamic quantities—such as entropy density $s$, energy density $\varepsilon$, the trace anomaly $(\varepsilon - 3P)$, specific heat $C_V$, and the speed of sound $c_s^2$—can be derived using standard thermodynamic relations. In our implementation, we include all hadronic states (up to $3~\mathrm{GeV}$) listed in the PDG2020 \cite{ParticleDataGroup:2020ssz} compilation, but exclude light nuclei as well as charm and bottom hadrons, whose contributions to the bulk thermodynamics at the relevant temperatures are negligible.
	
	
\section{Analysis Methodology: Bayesian Approach}
	\label{sec:analysis_methodology}
	We adopt a Bayesian framework to calibrate the parameters of the Cross EV--HRG model. In contrast to conventional $\chi^2$ minimization, which yields a single best-fit point with approximate Gaussian uncertainties, this approach provides the full posterior distribution in parameter space. This enables a consistent treatment of non-Gaussian uncertainties, parameter correlations, and degeneracies, while naturally incorporating priors and nuisance parameters within a unified probabilistic formulation. In practice, Markov Chain Monte Carlo (MCMC) sampling explores the global structure of the likelihood and is less susceptible to local minima than gradient-based optimization. When combined with surrogate models such as Gaussian Process (GP) emulators, this methodology allows for efficient inference with computationally demanding forward models and has become standard in contemporary Bayesian analyses of heavy-ion collisions~\cite{Wesolowski:2015fqa,Bernhard:2016tnd,Bernhard:2019bmu,Nijs:2020roc,JETSCAPE:2021ehl,Parkkila:2021yha}.
		
	In this work, the Cross EV--HRG model is calibrated without imposing any predefined functional form for the hadronic eigenvolumes. Instead, the effective hadron radii are treated as free parameters grouped according to baryon number and strangeness content. This flavor-based grouping is motivated by observed differences in interaction strengths and mass hierarchies among hadrons, as suggested by thermal-model studies employing LQCD thermodynamics and hadron yield data~\cite{Alba:2016hwx,Alba:2017bbr,Vovchenko:2016ebv}. Within the Bayesian framework, these parameters are constrained directly by confronting model predictions with both LQCD observables and experimental hadron yields.
	
 	The detailed methodology—including model implementation, data preprocessing, likelihood construction, and MCMC sampling—is presented in the following sections.
 		
 \subsection{Gaussian Process (GP) emulator}
	\label{subsec:GPR}
	The Cross EV--HRG model is computationally demanding because it requires solving a large system of coupled nonlinear equations—one for each hadron species—at every point in the parameter space. This complexity, combined with the high dimensionality of our flavor-dependent radii and freeze-out parameters, renders direct MCMC sampling using the full iterative solver impractical. To overcome this challenge, we construct a Gaussian process (GP) emulator using the \texttt{scikit-learn} library~\cite{Rasmussen:2006gpr,Pedregosa:2011sklearn}. The emulator serves as a surrogate for the full Cross EV--HRG calculation, interpolating its predictions across the parameter space and thereby significantly accelerating Bayesian sampling while preserving the accuracy required for parameter calibration.
	
 	\subsubsection{Training data sets}
	Separate emulators are trained for the lattice and yield analyses to accommodate their differing output structures and input dimensionalities. Training points are generated using Latin Hypercube Sampling (LHS)~\cite{McKay1979}, which ensures a quasi-random and well-distributed coverage of the multidimensional parameter space. The sampling ranges are listed in Table~\ref{tab:params}. The first eight parameters (species-dependent eigenvolume radii) enter both the lattice and yield analyses, whereas the chemical freeze-out temperature is included only in the yield emulator\textit{(The system radius is not included directly in GP training; number densities are rescaled at the MCMC stage)}.
	
	For the lattice analysis, 300 design points are sampled in an 8-dimensional eigenvolume radii space, while for the yield analysis, another 300 points are generated in a 9-dimensional space that also includes the freeze-out temperature. This choice of sample size was validated to provide sufficient GP accuracy at manageable computational cost; further details on the GP performance and validation are provided in the Supplementary Material. 	
 	
	\begin{table}[h!]
		\centering
		\begin{tabular}{|c|c|c|}
			\hline
			Parameter & Symbol & Range \\
			\hline
			non-strange meson radius & $r_{M}$ & 0.001--0.6 fm \\
			non-strange baryon radius & $r_{B}$ & 0.005--0.6 fm \\
			non-strange antibaryon radius & $r_{\bar{B}}$ & 0.005--0.6 fm \\
			singly strange meson radius & $r_{\mathrm{MS1}}$ & 0.005--0.6 fm \\
			doubly strange meson radius & $r_{\mathrm{MS2}}$ & 0.005--0.6 fm \\
			singly strange baryon radius & $r_{\mathrm{BS1}}$ & 0.005--0.6 fm \\
			doubly strange baryon radius & $r_{\mathrm{BS2}}$ & 0.005--0.6 fm \\
			triply strange baryon radius & $r_{\mathrm{BS3}}$ & 0.005--0.6 fm \\
			chemical freeze-out temperature & $T_{\mathrm{ch}}$ & 120--170 MeV \\
			system radius & $R_{\mathrm{ch}}$ & 2.2--14.0 fm \\
			\hline
		\end{tabular}
		\caption{Prior ranges of parameters used in the Cross EV--HRG model training sets. The first eight parameters listed in the table are common to both lattice and yield data, while chemical freeze-out temperature and system radius are included only in the yield calibration of a given centrality class.}
		\label{tab:params}
	\end{table}
 	
	We explicitly evaluate the Cross EV--HRG model at each training point to
	generate the corresponding outputs. For the lattice emulator, six thermodynamic
	observables are computed over temperatures between $110$ and $160~\text{MeV}$ in
	$5~\text{MeV}$ intervals, yielding 66 observables per design point and forming an
	output matrix $Y^{\text{lqcd}}_{300\times 66}$. For the yield analysis, the
	model computes eleven particle species, including all resonance-decay
	contributions. Considering all nine ALICE centrality classes, this results in an
	output matrix $Y^{\text{yield}}_{300\times 99}$.
		
	\subsubsection{PCA-Based Emulator Training and Validation}
	
	Because GP emulators operate on scalar outputs, we reduce the dimensionality of
	the model-output matrices using \textit{Principal Component Analysis} (PCA).
	PCA maps the correlated outputs onto an orthogonal basis of principal components
	(PCs).
	\[
	Z = YU,
	\]
	where \(Y\) is the standardized output matrix and \(U\), known as the transformation matrix, contains the eigenvectors of the covariance matrix \(Y^{T}Y\).
	
	The matrix \(U\) is obtained via a Singular Value Decomposition (SVD) of the form
	\(Y = V \Sigma U^{T}\), which implies
	\(Y^{T}Y = U \Sigma^{2} U^{T}\). Thus, the columns of \(U\) are the covariance
	eigenvectors, and the diagonal entries of \(\Sigma\) provide the singular values of \(Y\).
	
	The PCs are uncorrelated and ordered by decreasing explained variance. Retaining
	only the dominant \(q\) components yields an efficient reduced representation,
	\[
	Z_{q} = YU_{q}.
	\]
	In this work, 15 principal components are retained for the lattice observables
	and 11 for the yield observables, capturing more than 99.9\% of the total
	variance in both cases. These reduced components serve as the GP training targets and provide
	a substantial reduction in computational cost without compromising accuracy.
	Additional tests confirm that including more PCs does not further improve the
	emulator performance within the parameter ranges considered.
 	
	Each retained PC is modeled independently using a GP with an
	anisotropic squared–exponential kernel,
	\begin{equation}
		k(x,x')=\sigma_{\text{GP}}^{2}
		\exp\!\left[-\sum_{k}\frac{(x_{k}-x_{k}')^{2}}{2\,l_{k}^{2}}\right]
		+ \sigma_{n}^{2}\,\delta_{xx'},
	\end{equation}
	where $\sigma_{\text{GP}}^{2}$ is the signal variance, $l_{k}$ are
	parameter-specific length scales implementing automatic relevance determination (ARD), and $\sigma_{n}^{2}$ is a small noise term that absorbs residual numerical fluctuations and ensures a stable inversion of the covariance matrix. The hyperparameters are obtained by maximizing the
	marginal likelihood. 
	\begin{figure*}[t]
		\includegraphics[scale=0.4]{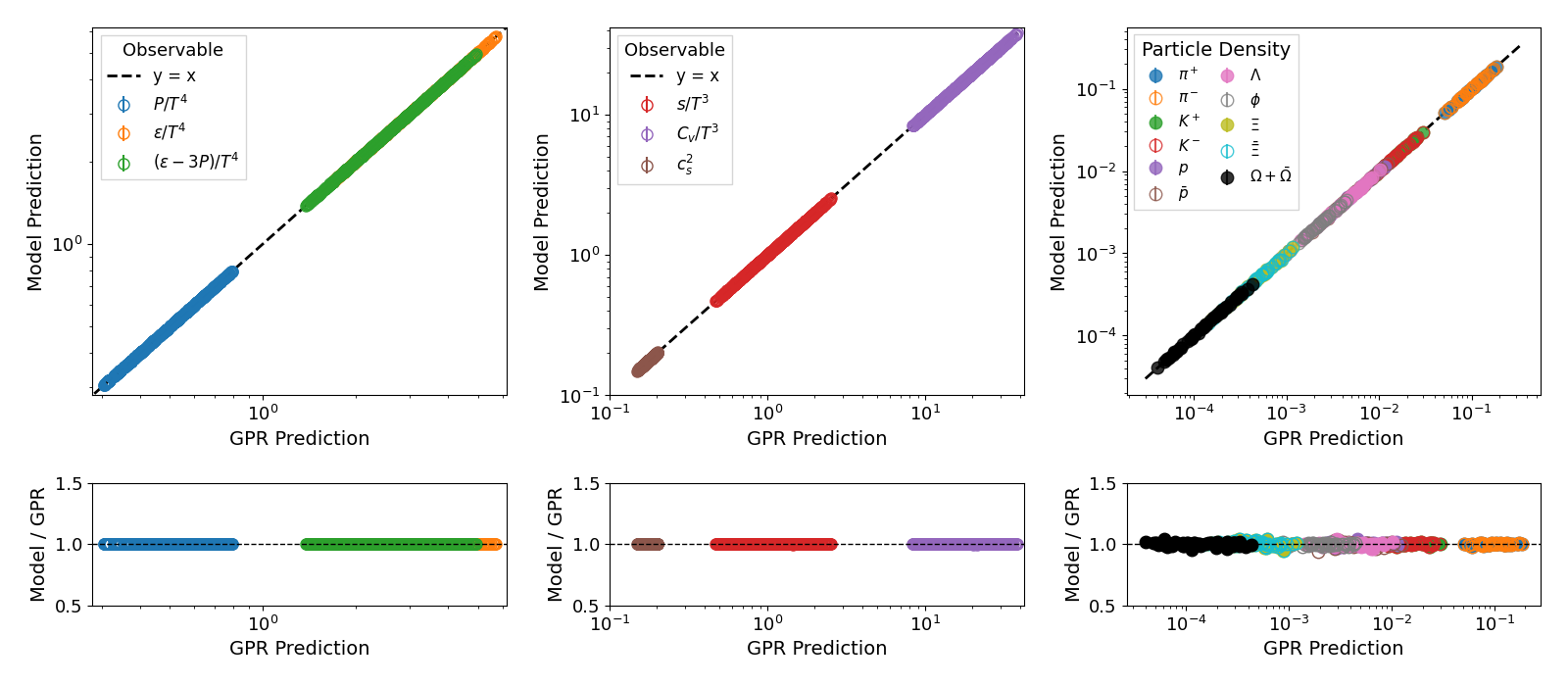}
		\caption{
			Validation of the GP emulator using 50 randomly
			selected test parameter sets. The legend identifies the thermodynamic and
			particle-density observables. The lower panel shows the ratio of full-model
			predictions to the emulator output, demonstrating the satisfactory accuracy of the GP
			approximation across all observables.
		}
		\label{fig:GPR_validation}
	\end{figure*}
   The trained emulators are validated using a fixed set of 50 randomly selected test points that are strictly excluded from all training sets. At each test point, the full set of observables entering the Bayesian calibration is evaluated using both the Cross EV--HRG model and the corresponding emulator. Excellent agreement between the model predictions and emulator outputs is observed across the explored parameter space, as illustrated in Fig.~\ref{fig:GPR_validation}. Detailed quantitative diagnostic metrics, including residual analyses and PCA truncation diagnostics, are provided in the Supplementary Material. These results confirm that the emulators achieve sufficient accuracy and provide reliable uncertainty estimates for use in the Bayesian analysis.

	\subsection{Bayesian Calibration}
	\label{subsec:bayesian_calibration}
	
	After verifying the predictive accuracy of the emulator, we proceed with the Bayesian calibration of the Cross EV--HRG model parameters through comparison with the model predictions with experimental hadron yield data and LQCD observables. This framework provides the full posterior distribution of the model parameters and naturally incorporates prior information, correlated uncertainties, and parameter degeneracies. Sampling of the posterior is carried out using MCMC, with the GP emulator enabling efficient evaluation of model predictions during the sampling procedure.
		
	Bayesian inference is based on
	\begin{equation}
		P(\theta | \mathcal{D}) \propto P(\mathcal{D} | \theta) P(\theta),
	\end{equation}
	where $P(\theta)$ denotes the prior and $P(\mathcal{D}|\theta)$ the likelihood. Here, $\theta$ denotes the model parameters and $\mathcal{D}$ the combined dataset.
	
	We adopt uniform (flat) priors within the design ranges listed in Table~\ref{tab:params} and zero probability outside:
	\begin{equation}
		P(\theta)\propto
		\begin{cases}
			1, & \min(\theta_i)\le\theta_i\le\max(\theta_i)\ \forall i,\\[2pt]
			0, & \text{otherwise}.
		\end{cases}
	\end{equation}
	This choice reflects the absence of additional physical prior information beyond the constraints encoded in the design ranges and avoids introducing unwarranted assumptions. For the LQCD calibration, we use only the eight eigenvolume parameters; for the yield calibration, two additional parameters—the chemical freeze-out temperature and the fireball radius for each centrality class—are included. The temperature prior is taken to be common across all centrality classes, while centrality-specific prior ranges are assigned for the fireball radius to facilitate fast MCMC convergence. In earlier EV--HRG studies, the extracted chemical freeze-out temperature often exhibited large uncertainties and a double-minimum $\chi^{2}$ profile~\cite{Vovchenko:2015cbk,Vovchenko:2016ebv}, with one minimum corresponding to an unphysical high-$T$ solution. As argued in Ref.~\cite{Vovchenko:2016ebv}, such high-temperature minima arise from the model’s sensitivity to the treatment of eigenvolume interactions—rather than reflecting genuine freeze-out conditions—because hadronic degrees of freedom are not expected to survive at those temperatures; therefore, in the present work we restrict the temperature prior to the pseudo-critical region.
 	
	The likelihood used in the calibration is
	\begin{equation}
		\label{Eq:likelihood}
		P(\mathcal{D}|\theta) \propto 
		\exp\!\left[-\tfrac{1}{2}\,(y_m(\theta)-y_e)^{T}
		\Sigma_t^{-1}\,(y_m(\theta)-y_e)\right],
	\end{equation}
	
	where $y_m(\theta)$ denotes the model predictions for parameter set $\theta$ and $y_e$ the corresponding experimental or LQCD data \textit{(All lattice and yield observables are scaled by their respective central values prior to entering the likelihood, rendering them dimensionless while preserving relative uncertainties; the effective weighting of the datasets is therefore governed by their covariance matrices rather than by absolute numerical scales)}. The total covariance matrix $\Sigma_t$ incorporates both the experimental/LQCD uncertainties and the emulator interpolation error. For each dataset, the GP predictive covariance is reconstructed by transforming the emulator variance from principal-component space back into observable space, after which the \textit{full GP covariance matrix} is added to the corresponding data--covariance matrix ($\Sigma_{\mathrm{lqcd}}$ or $\Sigma_{\mathrm{exp}}$), following Ref.~\cite{Bernhard:2018hnz}. We also include the additional calibration--uncertainty term $\sigma^{y}_{m}$ introduced in Ref.~\cite{Bernhard:2018hnz}, which accounts for the ``uncertainty of the uncertainty.'' 

	The joint likelihood for the simultaneous calibration of multiple datasets—in our case, LQCD observables and experimental hadron yields—can be written as the product of the individual likelihoods~\cite{Bernhard:2018hnz}:
	\begin{equation}
		\label{Eq:joint_likelihood}
		P(\mathcal{D}|\theta) = \prod_{s} P(\mathcal{D}_{s}|\theta).
	\end{equation}
	
	In the joint analysis, the lattice and yield likelihoods are characterized by distinct scales and uncertainty structures. We therefore omit the additional calibration-uncertainty term $\sigma^{y}_{m}$ in the combined fit, as a single shared parameter would not be well defined across the two datasets and could lead to an imbalanced weighting of their respective contributions to the joint likelihood.
	
    For all calibrations, we sample the posterior distribution using the affine-invariant ensemble Markov Chain Monte Carlo (MCMC) sampler~\cite{ForemanMackey2013}, as implemented in the \texttt{emcee} package. An initial burn-in phase of $\mathcal{O}(10^{5})$ steps is discarded after equilibration, followed by $\mathcal{O}(10^{6})$ production samples used for posterior estimation. To improve convergence, the burn-in procedure is performed in two stages.
    
	Convergence is assessed using both the rank-normalized and split potential scale reduction statistics~\cite{Vehtari:2021ysn}, with $\hat R_{\mathrm{rn}}$ and $\hat R_{\mathrm{split}} \approx 1$ for all parameters. The bulk effective sample size satisfies $\mathrm{ESS}_{\mathrm{bulk}} > 400$, as recommended in Ref.~\cite{Vehtari:2021ysn}. In addition, the autocorrelation-based effective sample size~\cite{GoodmanWeare2010} exceeds $\mathrm{ESS}_{\mathrm{int}} \gtrsim 1300$, ensuring reliable posterior estimation with Monte Carlo uncertainties remaining at the few-percent level.
	
	All corner plots shown in this work are constructed from the converged post–burn-in ensemble. The marginalized 1D and 2D posteriors, their credible intervals contours, and all MAP values and summary statistics are computed from this converged ensemble. The full analysis code is available in Ref.~\cite{HepBayes:2023}.
 	
	\section{Results and Discussion}
	\label{results_discussion}
 	\begin{figure*} 
		\includegraphics[scale=0.50]{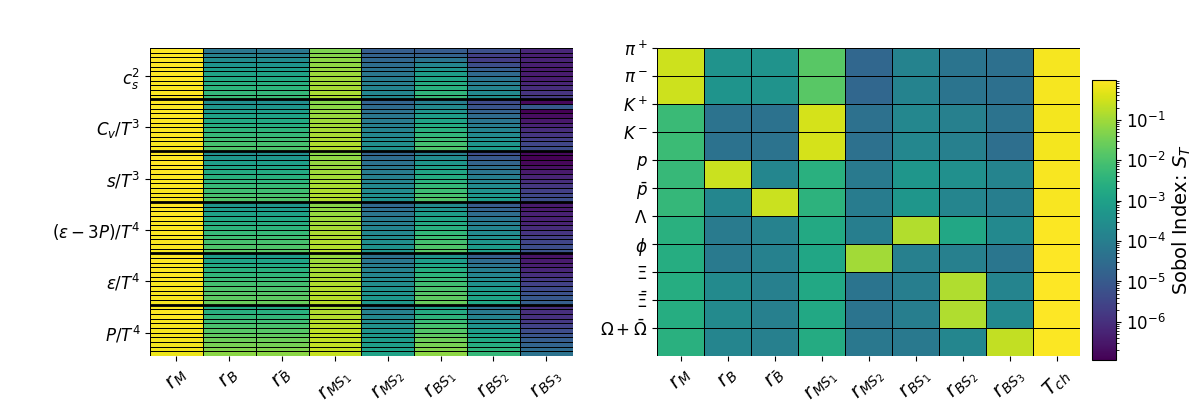}
		
		\caption{
			Total Sobol index ($S_T$) heatmaps illustrating the global sensitivity of model parameters to different observables. \textbf{Left panel:} Sensitivity of LQCD thermodynamic observables to species-dependent eigenvolume parameters. Each row corresponds to an observable, and each column to a specific eigenvolume parameter; horizontal blocks represent different temperature points in the LQCD dataset, ordered from low to high temperature. \textbf{Right panel:} Sensitivity of hadron yield observables to the model parameters, including the freeze-out temperature $T_{\mathrm{ch}}$. The color scale represents the total Sobol index on a logarithmic scale.
		}
 		\label{fig:sobol_plot}
	  \end{figure*}
  	
	\subsection{Sensitivity analysis }
	
 	We begin the results section by assessing the global sensitivity of the model parameters using the total Sobol indices ($S_T$)~\cite{Sobol2001}. Separate analyses are performed for LQCD thermodynamic observables and for experimental hadron yields, as shown in Fig.~\ref{fig:sobol_plot}. The corresponding
	heatmaps quantify the contribution of each model parameter to each observable, as detailed in the figure caption.
 	
	For the LQCD observables (Fig.~\ref{fig:sobol_plot}, left panel), the sensitivity is dominated by the eigenvolume parameters associated with the most abundant species at $\mu_B \approx 0$, namely the non-strange meson radius $r_M$ and the singly-strange meson radius $r_{MS_1}$. Conversely, the triply-strange baryon radius $r_{BS_3}$ exhibits negligible influence, consistent with the small abundance and limited thermodynamic weight of such states. The total Sobol indices show no significant temperature dependence for any of the thermodynamic observables, indicating that the relative importance of the model parameters remains approximately constant over the considered temperature range.
 	For the hadron yields (Fig.~\ref{fig:sobol_plot}, right panel), the sensitivity pattern becomes strongly species dependent. The chemical freeze-out temperature, $T_{\mathrm{ch}}$ emerges as the most influential parameter across all yields, reflecting its central role in setting the overall abundance scale. As expected, the sensitivity of eigenvolume parameters contribute in a flavor-dependent manner that closely tracks the quark content of each hadron species.
	
	\subsection{MCMC calibration results}
	As discussed earlier, we calibrate the Cross EV--HRG model to LQCD thermodynamic observables and ALICE hadron yields using a Bayesian framework. To assess the relative constraining power of each dataset and the impact of correlated experimental systematics, we perform four independent calibrations: (i) LQCD-only; (ii) yields with a diagonal experimental covariance; (iii) yields with a full phenomenological covariance including intra-species and centrality correlations; and (iv) a joint LQCD+yield calibration employing the full covariance of systematic uncertainties for yields.	
	
 	\begin{figure}[t]
		\includegraphics[scale=0.20]{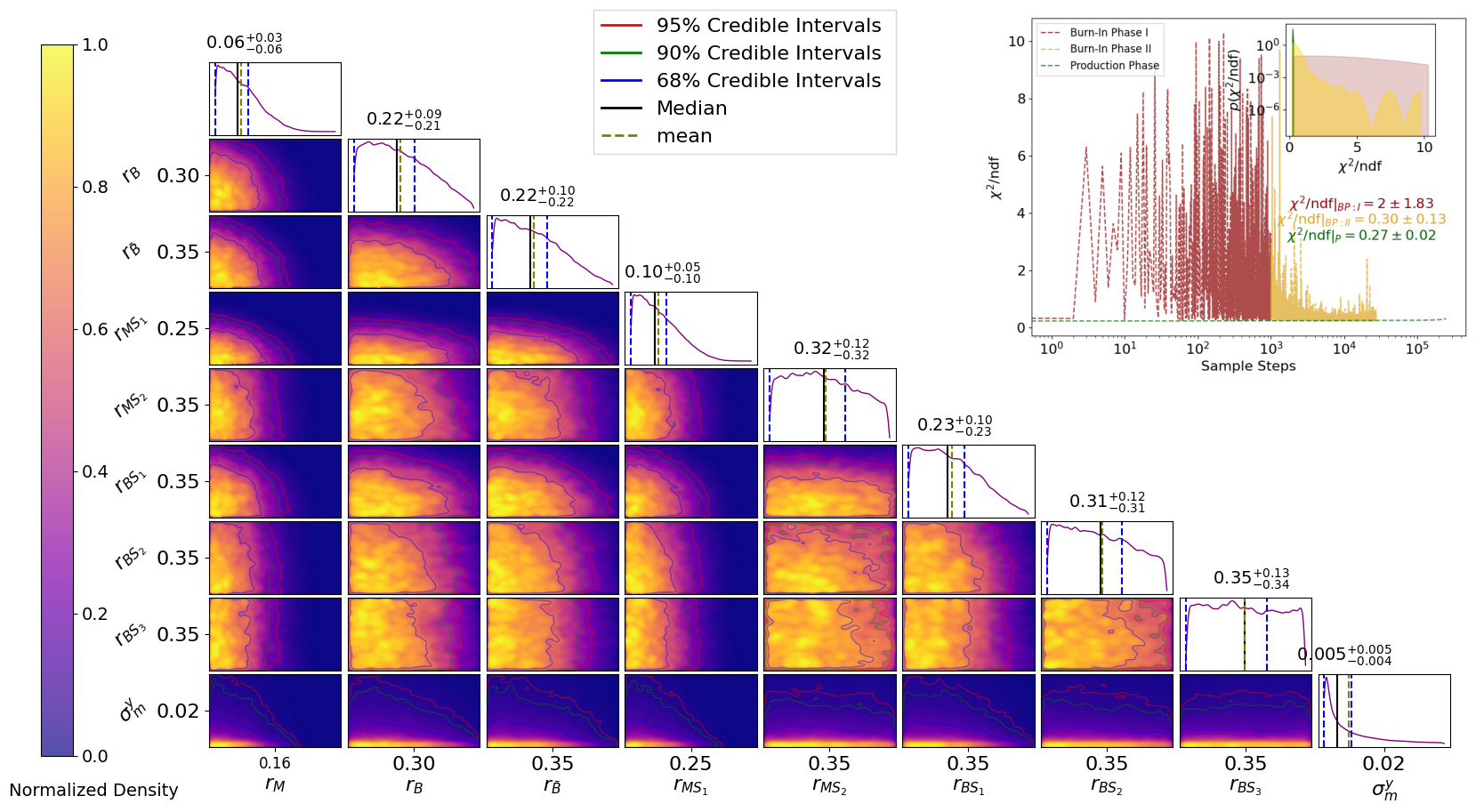}		
		\caption{ Corner plot showing the posterior distributions and pairwise correlations of the model parameters obtained from the Bayesian calibration of the Cross EV--HRG model to LQCD data. The diagonal panels display the one-dimensional marginalized posteriors, with the posterior median, mean, and 68\% credible intervals indicated in the legend. The off-diagonal panels show the corresponding two-dimensional joint posterior distributions with credible contours at the indicated levels. The color bar represents the normalized posterior density. The inset shows the evolution of $\chi^2/\mathrm{ndf}$ during the MCMC sampling, including the burn-in phase,
			along with the marginalized $\chi^2/\mathrm{ndf}$ distribution.
		}
		\label{fig:corner_plot_lqcd}
	\end{figure}
 	\begin{figure*}
		\includegraphics[scale=0.4]{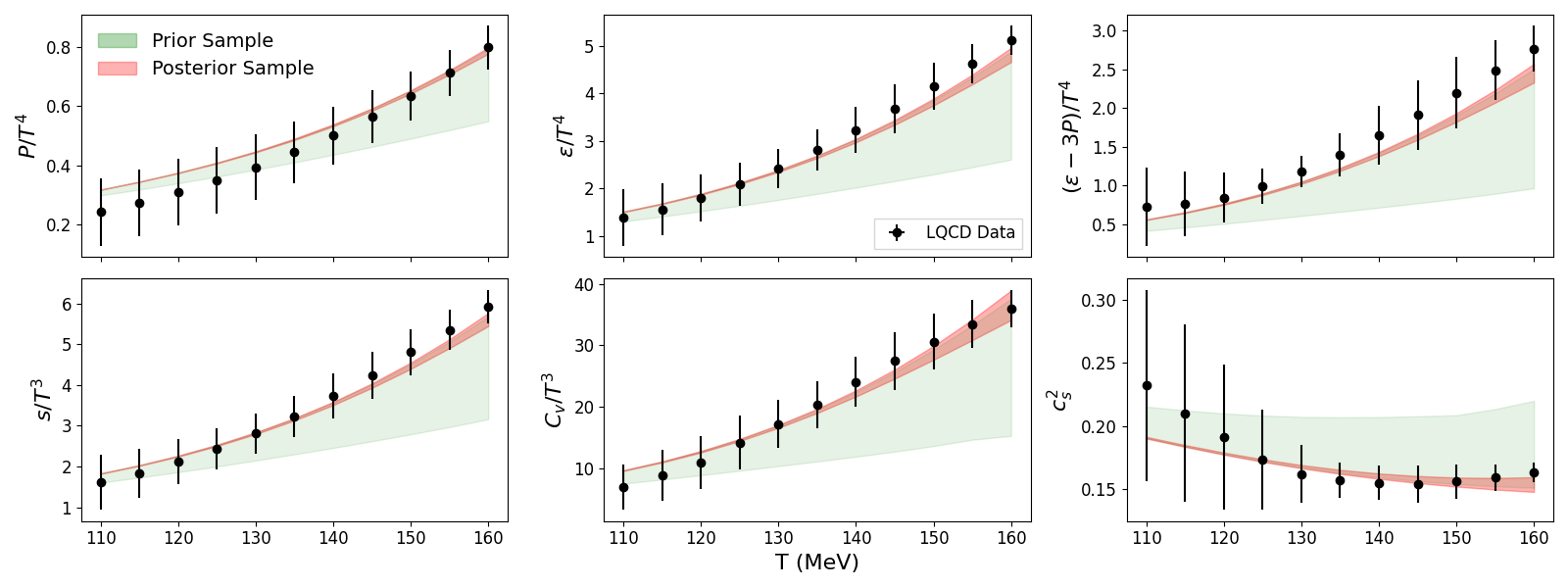}		
		\caption{
			Comparison between LQCD results for selected thermodynamic observables and model predictions from the Cross EV--HRG calibration. The broad green band represents model samples drawn from the pre-converged burn-in phase, while the narrower red band corresponds to predictions obtained from the marginalized posterior distribution.
		}
 		\label{fig:posterior_validation_plot_lqcd_only}
	\end{figure*}

 	\subsubsection{Lattice Data Fitting}
	\label{subsub:lqcd_only}
	
	For the calibration using lattice data, we consider six key thermodynamic observables: normalized pressure ($P/T^4$), energy density ($\varepsilon/T^4$), entropy density ($s/T^3$), specific heat ($C_V/T^3$), trace anomaly [$(\varepsilon - 3P)/T^4$], and the squared speed of sound ($c_s^2$). These observables are taken from the LQCD results reported in Ref.~\cite{Borsanyi:2013bia}. The methodology and implementation details of the GP framework and Bayesian calibration are provided in Sec.~\ref{sec:analysis_methodology}. The main results of the LQCD calibration are presented in Figs.~\ref{fig:corner_plot_lqcd} and~\ref{fig:posterior_validation_plot_lqcd_only} (see the figure captions for details), with corresponding numerical values summarized in Table~\ref{table:mcmc_lqcd_only}. 
	The inset of Fig.~\ref{fig:corner_plot_lqcd} shows the evolution of \( \chi^2/\mathrm{ndf} \) during the burn-in and production phases. The chi-squared statistic is defined as
	\begin{equation}
		\chi^{2} =
		\bigl(\mathbf{y}_{\rm model} - \mathbf{y}_{\rm lqcd}\bigr)^{\!\top}
		\mathbf{\Sigma}_{t}^{-1}
		\bigl(\mathbf{y}_{\rm model} - \mathbf{y}_{\rm lqcd}\bigr),
	\end{equation}
	where $\mathbf{\Sigma}_{t}$ denotes the total covariance matrix, given by 
	$
	\mathbf{\Sigma}_{t}
	= \mathbf{\Sigma}_{\rm LQCD}^{\rm (diagonal)}
	+ \mathbf{\Sigma}_{\rm GP},
	$
	which incorporates both LQCD statistical uncertainties and emulator uncertainties. The number of degrees of freedom, $\mathrm{ndf}$, is defined as the number of independent observables entering the calibration minus the number of free model parameters. The evolution of \( \chi^2/\mathrm{ndf} \) serves not only as a measure of the goodness of fit but also as an additional diagnostic for the convergence of the MCMC sampler. A stable and decreasing trend in the chi-squared values typically indicates that the chains have equilibrated and are sampling from the high-probability regions of the posterior distribution.
 	
	A visual validation of the calibrated model is shown in Fig.~\ref{fig:posterior_validation_plot_lqcd_only}. The broad green bands, generated from burn-in samples, demonstrate that the chosen priors adequately span the relevant parameter space. 
	The narrower red bands, obtained from the converged chains, represent the posterior predictions and closely follow the LQCD data, indicating that the calibrated model provides a good description of the thermodynamic observables. The overall agreement is quantified by a reduced chi-squared value of \( \chi^2/\mathrm{ndf} \sim 0.27 \pm 0.02 \).
	
 	\begin{table}[h!]
		\centering
		\resizebox{1.05\linewidth}{!}{%
			\begin{tabular}{@{}ccccccc@{}} 
				\toprule
				\textbf{Par} & \textbf{Mean} & \textbf{Median} & \textbf{MAP} & \textbf{CI (68\%)} & \textbf{CI (90\%)} & \textbf{CI (95\%)} \\
				\midrule
				\textbf{$r_{M}$}       
				& 0.069 & 0.060 & 0.013 & 0.019--0.120 & 0.006--0.164 & 0.004--0.183 \\
				
				\textbf{$r_{B}$}       
				& 0.234 & 0.216 & 0.234 & 0.070--0.405 & 0.023--0.508 & 0.012--0.545 \\
				
				\textbf{$r_{\bar{B}}$}      
				& 0.245 & 0.223 & 0.244 & 0.067--0.427 & 0.021--0.551 & 0.011--0.602 \\
				
				\textbf{$r_{MS_{1}}$}      
				& 0.117 & 0.103 & 0.046 & 0.032--0.207 & 0.010--0.276 & 0.006--0.307 \\
				
				\textbf{$r_{MS_{2}}$}      
				& 0.332 & 0.321 & 0.171 & 0.108--0.563 & 0.037--0.654 & 0.020--0.677 \\
				
				\textbf{$r_{BS_{1}}$}      
				& 0.256 & 0.235 & 0.409 & 0.077--0.438 & 0.028--0.562 & 0.017--0.611 \\
				
				\textbf{$r_{BS_{2}}$}     
				& 0.326 & 0.314 & 0.374 & 0.099--0.560 & 0.033--0.654 & 0.019--0.677 \\
				
				\textbf{$r_{BS_{3}}$}     
				& 0.349 & 0.346 & 0.506 & 0.113--0.588 & 0.039--0.667 & 0.022--0.684 \\
				
				\textbf{$\sigma^y_m$}  
				& 0.005 & 0.005 & 0.001 & 0.002--0.018 & $<$0.030 &$<$0.034 \\
				\bottomrule
		\end{tabular}}
		\caption{
			Summary statistics of the posterior distributions for the model parameters obtained from the MCMC calibration of the Cross EV-HRG model using LQCD data only. Shown are the posterior mean, median, and maximum a posteriori (MAP) estimates, together with the 68\%, 90\%, and 95\% credible intervals. All radii are given in femtometers (fm). The parameter $\sigma^y_m$ denotes the calibration-uncertainty term. 			
		}
 		\label{table:mcmc_lqcd_only}
	\end{table}	
	

	The posterior medians of the non-strange baryon and antibaryon radii are statistically indistinguishable, consistent with baryon–antibaryon symmetry at vanishing baryon chemical potential ($\mu_B = 0$). The strange baryonic radii exhibit a monotonic ordering in their posterior median and mean values, $r_{BS_1} < r_{BS_2} < r_{BS_3}$. This hierarchy deviates from the commonly assumed inverse mass-scaling behavior for strange hadrons, wherein heavier states are expected to possess smaller eigenvolumes~\cite{Alba:2016hwx}. However, given the substantial and overlapping uncertainties, no definitive conclusion can be drawn regarding the physical significance of this ordering. The extracted radii for non-strange and single-strange mesons are consistent with zero within uncertainties. Both posterior distributions peak near zero, indicating that light mesons—particularly pions and kaons—behave as quasi-pointlike particles with negligible short-range repulsive interactions. This observation provides a statistical justification for the “baryon-only” exclusion schemes frequently employed in the literature~\cite{Andronic:2012ut,Kadam:2019rzo}.
	
	Most importantly, as shown in Table~\ref{table:mcmc_lqcd_only}, the eigenvolume parameters are only weakly constrained by LQCD data alone. Several parameters exhibit large credible intervals, with noticeable separations between the MAP values and the posterior means and medians, indicating weakly informative posteriors. This behavior arises because bulk thermodynamic quantities, such as the pressure, energy density, and trace anomaly, are integrals over the full hadronic spectrum and are therefore only weakly sensitive to species dependent eigenvolume effects, a conclusion that is also supported by the Sobol analysis shown in Fig.~\ref{fig:sobol_plot}. As a consequence, earlier studies were able to reproduce lattice thermodynamics even when employing a common eigenvolume for all hadrons~\cite{Sarkar:2017ijd,Vovchenko:2014pka}. Notably, these wide uncertainties persist despite achieving a low $\chi^2/\mathrm{ndf}$, illustrating that a good quality-of-fit to bulk thermodynamics does not necessarily imply well-identified individual parameters.
	
	Part of this limited sensitivity, particularly in the strange sector, likely reflects the incomplete inclusion of higher-mass hadronic states in the PDG spectrum. In the mesonic sector, the near-zero value of $r_M$ may similarly signal incompleteness of the meson spectrum, as the absence of heavier resonances reduces the effective number of active degrees of freedom and biases the fit toward smaller eigenvolumes—an effect consistent with studies incorporating quark-model states~\cite{Sarkar:2023cnq,Alba:2017mqu, Alba:2017bbr,Karthein:2021cmb,Bazavov:2014xya} or Hagedorn-type extensions \cite{Sarkar:2017ijd,Vovchenko:2014pka}. Quark-model calculations predict a significantly richer hadronic spectrum~\cite{Ebert:2009ub}, the inclusion of which would be expected to enhance sensitivity to flavor-dependent eigenvolume effects. Beyond this, complementary observables outside bulk thermodynamics—such as experimental hadron yields or higher-order conserved-charge fluctuations—are essential for resolving the remaining parameter degeneracies and improving identifiability.
 	
	\begin{figure*}[t]
		\includegraphics[scale=0.40]{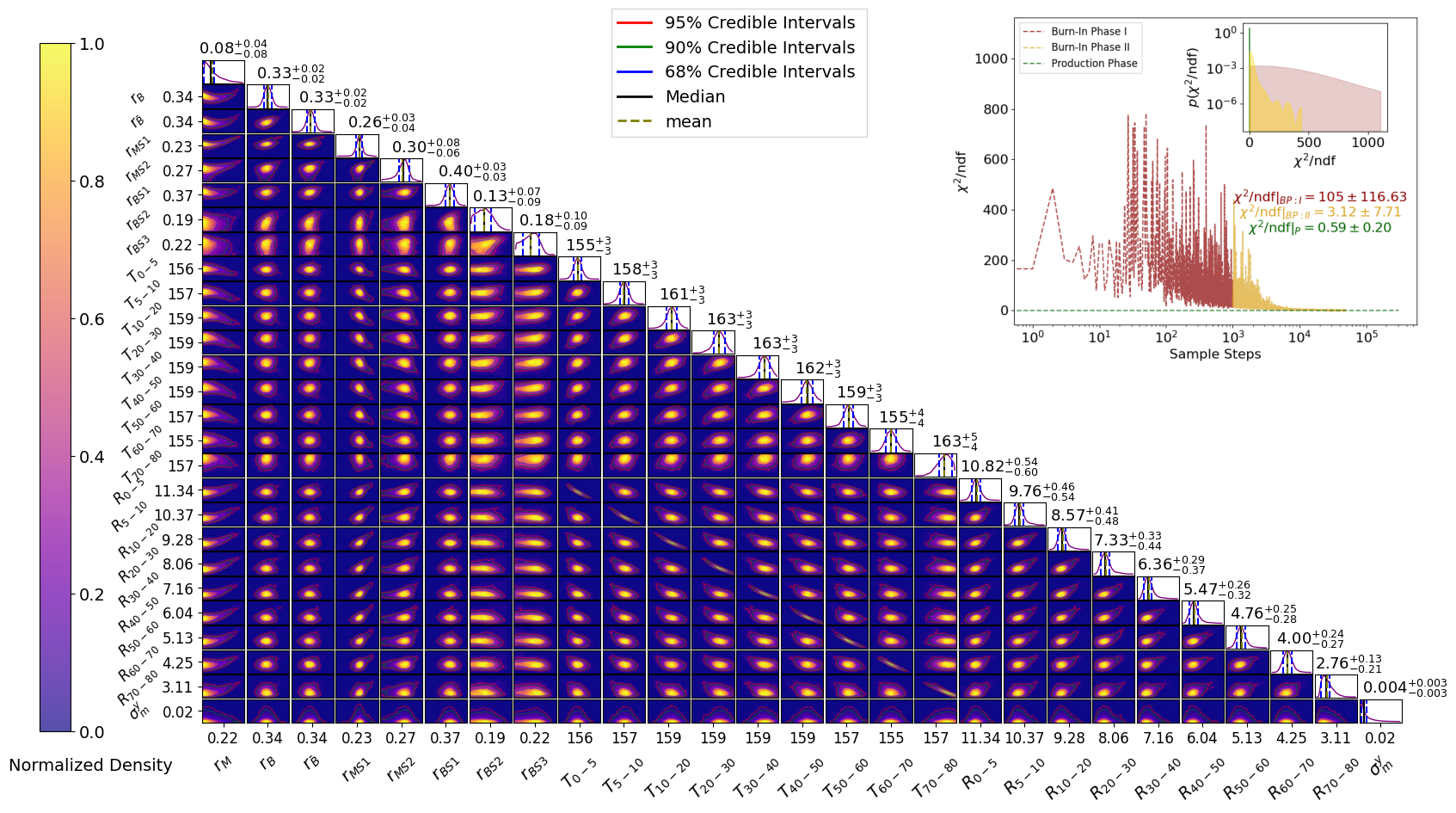}
 		\caption{
 			Same as the caption of Fig.~\ref{fig:corner_plot_lqcd}, except that the Bayesian fit is performed using centrality-dependent hadron yield data, with experimental uncertainties treated via a diagonal covariance matrix.
 		}
 		
		\label{fig:corner_plot_only_yield_diagonal_cor}
	\end{figure*}
	\begin{table}[htbp]
		\scriptsize
		\centering
		\renewcommand{\arraystretch}{1.5} 
		\resizebox{1.05\linewidth}{!}{%
			\begin{tabular}{lcccccc}
				\toprule
				\small
				\textbf{Parameter} & \textbf{Mean} & \textbf{Median} & \textbf{MAP} &
				\textbf{68\% CI} & \textbf{90\% CI} & \textbf{95\% CI} \\
				\midrule
				$r_M$       & 0.101 & 0.077 & 0.187 & 0.022 -- 0.187 & 0.007 -- 0.273 & 0.004 -- 0.314 \\
				$r_B$       & 0.332 & 0.332 & 0.323 & 0.310 -- 0.354 & 0.296 -- 0.370 & 0.288 -- 0.379 \\
				$r_{\bar B}$& 0.329 & 0.329 & 0.327 & 0.307 -- 0.351 & 0.292 -- 0.366 & 0.284 -- 0.374 \\
				$r_{MS_1}$  & 0.262 & 0.262 & 0.284 & 0.227 -- 0.296 & 0.198 -- 0.331 & 0.177 -- 0.358 \\
				$r_{MS_2}$  & 0.287 & 0.300 & 0.335 & 0.213 -- 0.363 & 0.112 -- 0.402 & 0.065 -- 0.425 \\
				$r_{BS_1}$  & 0.399 & 0.400 & 0.383 & 0.370 -- 0.428 & 0.350 -- 0.446 & 0.340 -- 0.455 \\
				$r_{BS_2}$  & 0.133 & 0.131 & 0.167 & 0.051 -- 0.210 & 0.021 -- 0.258 & 0.013 -- 0.281 \\
				$r_{BS_3}$  & 0.179 & 0.185 & 0.189 & 0.075 -- 0.272 & 0.028 -- 0.315 & 0.017 -- 0.332 \\
				\midrule
				$T_{0\text{--}5}$   & 155.4 & 155.4 & 154.5 & 152.6 -- 158.2 & 150.9 -- 160.3 & 150.0 -- 161.4 \\
				$R_{0\text{--}5}$   & 10.87 & 10.82 & 11.05 & 10.29 -- 11.43 & 9.93 -- 11.94 & 9.75 -- 12.23 \\
				\midrule
				$T_{5\text{--}10}$   & 158.0 & 158.0 & 156.4 & 155.2 -- 160.8 & 153.0 -- 162.6 & 151.8 -- 163.5 \\
				$R_{5\text{--}10}$   & 9.83 & 9.76 & 10.13 & 9.30 -- 10.32 & 9.03 -- 10.90 & 8.90 -- 11.27 \\
				\midrule
				$T_{10\text{--}20}$  & 160.6 & 160.6 & 160.7 & 157.7 -- 163.5 & 155.4 -- 165.4 & 154.2 -- 166.4 \\
				$R_{10\text{--}20}$   & 8.63 & 8.57 & 8.60 & 8.16 -- 9.07 & 7.91 -- 9.60 & 7.79 -- 9.90 \\
				\midrule
				$T_{20\text{--}30}$   & 162.6 & 162.7 & 160.9 & 159.7 -- 165.7 & 157.1 -- 167.6 & 155.8 -- 168.4 \\
				$R_{20\text{--}30}$  & 7.39 & 7.33 & 7.64 & 6.98 -- 7.76 & 6.78 -- 8.25 & 6.70 -- 8.55 \\
				\midrule
				$T_{30\text{--}40}$   & 162.8 & 162.9 & 160.8 & 159.7 -- 166.0 & 157.3 -- 167.8 & 155.8 -- 168.6 \\
				$R_{30\text{--}40}$  & 6.41 & 6.36 & 6.63 & 6.06 -- 6.74 & 5.89 -- 7.13 & 5.82 -- 7.41 \\
				\midrule
				$T_{40\text{--}50}$   & 162.0 & 162.0 & 161.4 & 158.9 -- 165.2 & 156.5 -- 167.3 & 155.2 -- 168.3 \\
				$R_{40\text{--}50}$   & 5.51 & 5.47 & 5.61 & 5.20 -- 5.79 & 5.04 -- 6.12 & 4.97 -- 6.35 \\
				\midrule
				$T_{50\text{--}60}$   & 158.7 & 158.6 & 159.8 & 155.5 -- 162.0 & 153.2 -- 164.3 & 151.9 -- 165.6 \\
				$R_{50\text{--}60}$   & 4.78 & 4.76 & 4.77 & 4.50 -- 5.04 & 4.35 -- 5.29 & 4.27 -- 5.47 \\
				\midrule
				$T_{60\text{--}70}$   & 155.3 & 155.2 & 154.4 & 151.4 -- 159.3 & 148.9 -- 162.1 & 147.8 -- 163.4 \\
				$R_{60\text{--}70}$   & 4.01 & 4.00 & 4.11 & 3.75 -- 4.26 & 3.60 -- 4.47 & 3.53 -- 4.58 \\
				\midrule
				$T_{70\text{--}80}$   & 162.7 & 163.2 & 162.9 & 158.1 -- 167.3 & 154.8 -- 169.0 & 153.2 -- 169.5 \\
				$R_{70\text{--}80}$   & 2.80 & 2.76 & 2.80 & 2.61 -- 2.99 & 2.55 -- 3.17 & 2.53 -- 3.26 \\
				\midrule
				$\sigma^y_m$ & 0.0043 & 0.0040 & 0.0026 & 0.0016 -- 0.0118 & 0.0012 -- 0.0200 & 0.0011 -- 0.0253 \\
				\bottomrule
			\end{tabular}			
		} 
		\caption{Summary of posterior statistics for various parameters obtained from centrality-dependent hadron yield data at $\sqrt{s_{\mathrm{NN}}} = 2.76$~TeV Pb--Pb collisions, neglecting correlations in the systematic uncertainties.
		}		
		\label{tab:mcmc_summary_yield_diagonal_cor}		
	\end{table}
	\begin{figure*}
		\includegraphics[scale=0.4]{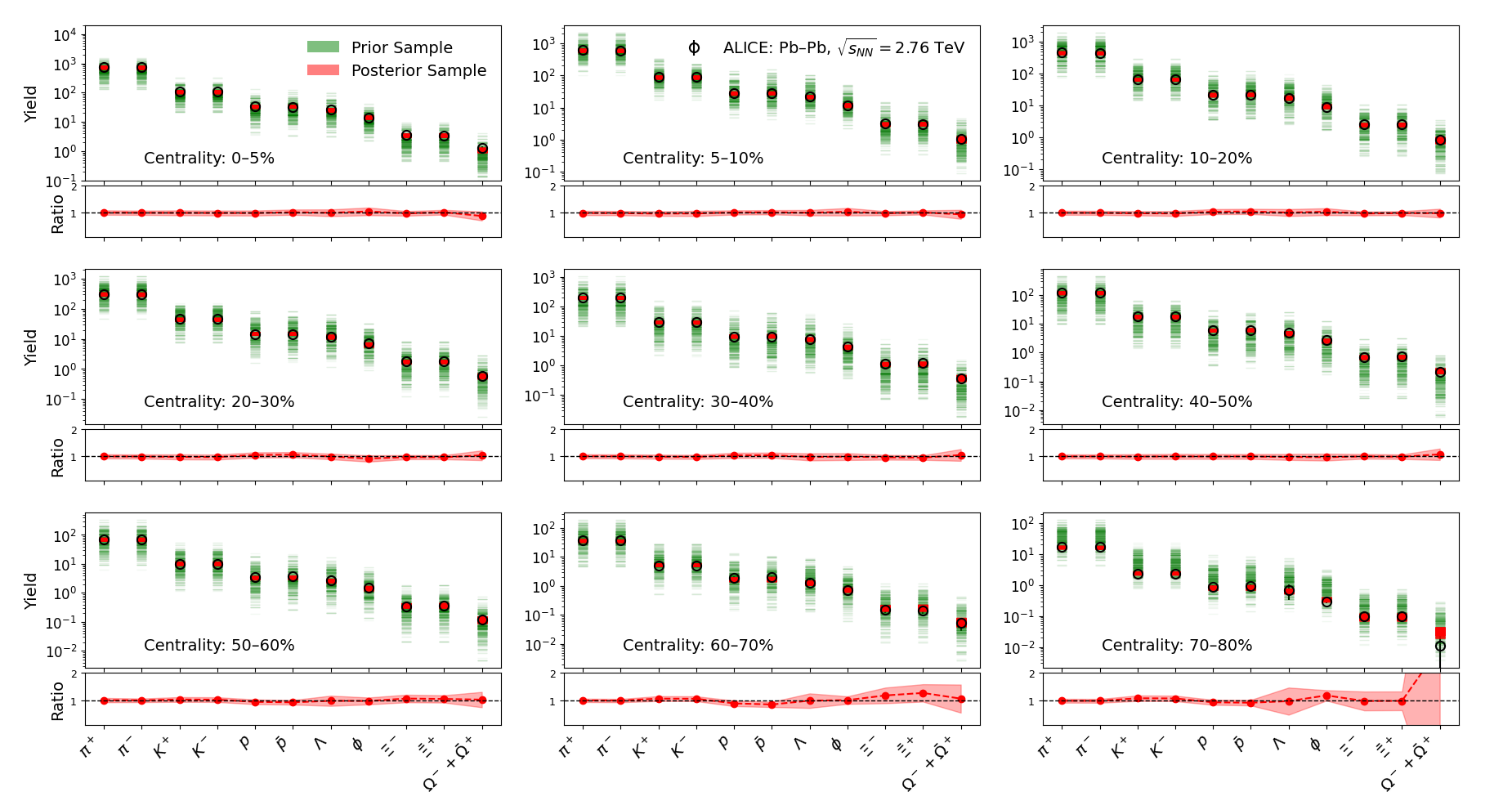}			
		\caption{Centrality-dependent comparison between the experimental hadron yields measured by ALICE at \(\sqrt{s_{NN}} = 2.76~\mathrm{TeV}\) and the model predictions. The broad green bands correspond to model predictions based on parameter sets drawn from the burn-in phase, while the narrow red bands represent predictions obtained using marginalized posterior samples.} 		%
		\label{fig:posterior_validation_plot_yield_diagonal}
	\end{figure*}

\subsubsection{Yield Data Fitting Without Correlations}
	\noindent
	
    We perform a Bayesian calibration using midrapidity hadron yields measured by the ALICE Collaboration in Pb--Pb collisions at $\sqrt{s_{\mathrm{NN}}}=2.76~\mathrm{TeV}$. The fit includes light hadrons ($\pi^{\pm}, K^{\pm}, p, \bar{p}$)~\cite{ALICE:2013mez}, singly-strange baryons ($\Lambda$)~\cite{ALICE:2013cdo}, hidden-strangeness mesons ($\phi$)~\cite{ALICE:2014jbq}, and multi-strange baryons ($\Xi^{\pm},\,\Omega+\bar{\Omega}$) across all available centrality classes~\cite{ALICE:2013xmt,Becattini:2014hla}. Experimental uncertainties are treated using a diagonal covariance matrix, with statistical and systematic errors added in quadrature and no correlations assumed between different particle species or centrality bins. The GP predictive covariance is incorporated following the LQCD calibration procedure, and only the experimental component of the total covariance matrix is diagonal. The impact of correlated systematic uncertainties is examined separately in the next section.
 	
	The yield-only calibration results are presented in Figs.~\ref{fig:corner_plot_only_yield_diagonal_cor} and~\ref{fig:posterior_validation_plot_yield_diagonal}, with numerical values listed in Table~\ref{tab:mcmc_summary_yield_diagonal_cor}.  
	
	The hadronic-yield calibration leads to a substantial tightening of the eigenvolume parameter posteriors relative to the LQCD-only calibration, typically at the level of $\sim60$--$90\%$, with the notable exception of the non-strange meson radius. Complementary statistical measures, such as the Jensen--Shannon divergence (JSD)~\cite{Endres2003JSDmetric}, with values in the range $\sim0.3$--$0.6$, indicate that this tightening is accompanied by a genuine reshaping of the posterior distributions rather than a simple uniform contraction. Among the eigenvolume parameters, those associated with the non-strange and singly-strange baryon sectors, as well as singly-strange mesons, are the most tightly constrained. Parameters associated with the double-strange and triple-strange sectors remain only moderately constrained, exhibiting broad and often asymmetric posterior distributions.	
	 
	The temperature and system radius parameters are likewise well constrained across centrality classes, with only mild broadening observed in the most peripheral bins, consistent with the larger experimental uncertainties in those data. This improvement reflects the strong sensitivity of hadron yields to excluded-volume effects and chemical freeze-out conditions,~\cite{Andronic:2017pug,Vovchenko:2016ebv,Vovchenko:2015cbk}, as also indicated by the Sobol sensitivity indices shown in Fig.~\ref{fig:sobol_plot}.   The comparison between model predictions and data in Fig.~\ref{fig:posterior_validation_plot_yield_diagonal} shows good agreement for the centrality-dependent yields, with a reduced chi-squared value of $\chi^2/\mathrm{ndf} \sim 0.59 \pm 0.20$, and only small deviations observed in peripheral collisions for multi-strange hadrons. 
	
	The extracted chemical freeze-out temperature exhibits only a mild variation across centrality classes and remains confined to a narrow range, $T_{\mathrm{ch}} \simeq 155$--$163~\mathrm{MeV}$, with the corresponding $68\%$ credible intervals spanning approximately $152$--$167~\mathrm{MeV}$. Within these uncertainties, no statistically significant centrality dependence is observed. This behavior is consistent with earlier analyses~\cite{Biswas:2020dsc} and is commonly interpreted as being compatible with approximate thermalization across all centrality classes~\cite{Kumar:2023acr,Gupta:2014ova}. As in the LQCD-only calibration, the yield-only fit allows for a small but finite non-strange meson radius. In particular, the posterior for $r_M$ has a mean value of $r_M \simeq 0.10~\mathrm{fm}$ with a broad credible interval, indicating that the meson sector remains only weakly constrained and is consistent with near point-like behavior within current uncertainties. The baryon and antibaryon radii remain nearly degenerate ($r_B \approx r_{\bar{B}}$), reflecting matter--antimatter symmetry at vanishing baryon chemical potential. The absence of a statistically significant splitting implies that any potential annihilation-related effects cannot be resolved within the present EV--HRG framework.

    A flavor-dependent structure emerges in the strange baryon sector; however, it does not exhibit a clear correspondence with the inverse mass hierarchy of repulsive interactions—where heavier particles are typically associated with smaller effective radii—previously suggested in the literature~\cite{Alba:2016hwx}. In particular, the extracted mean and median radii of double-strange baryons are found to be smaller than those of triple-strange baryons. This ordering must be interpreted with caution, given the sizable uncertainties and broad posterior distributions characterizing the multi-strange sectors. Contributions from sequential freeze-out effects~\cite{Chatterjee:2013yga,Bugaev:2013sfa,Chatterjee:2014ysa} may also play a role and cannot be excluded at the present level of precision. It is therefore plausible that residual uncertainties associated with the incompleteness of the PDG hadron spectrum and resonance feeddown~\cite{Andronic:2017pug,Vovchenko:2020dmv} could influence, and potentially clarify, the observed flavor-dependent structure in the strange baryon sector.
    
    
    The off-diagonal contours of Fig.~\ref{fig:corner_plot_only_yield_diagonal_cor} reveal a nontrivial correlation structure among the model parameters. Light-hadron radii show systematic positive correlations ($|\rho|\sim0.25$--$0.45$), with the strongest correlation observed between baryon and antibaryon radii ($\rho(r_B,r_{\bar B})\simeq0.43$), consistent with baryon--antibaryon symmetry at $\mu_B\simeq0$. Correlations involving multi-strange hadrons are progressively weaker, being moderate for double-strange species and small for triple-strange ones, reflecting the limited constraining power of current multi-strange hadron yield data.
    
    A strong temperature--volume degeneracy is observed, with $|\rho(T_i,R_i)|\gtrsim0.9$ across all centrality classes. Light-hadron eigenvolumes exhibit positive correlations with the system radius $R_{\mathrm{ch}}$ ($\rho\sim0.45$--$0.55$) and moderate anti-correlations with the temperature $T_{\mathrm{ch}}$ ($\rho\sim-0.35$ to $-0.45$), indicating that repulsive effects in the most abundant hadron species are predominantly accommodated through an effective expansion of the system radius. In contrast, less abundant species such as multi-strange hadrons display weaker and less systematic correlations with $R_{\mathrm{ch}}$ ($|\rho|\lesssim0.3$), as variations of the system radius are tightly constrained by the light-hadron sector and are therefore disfavored as a mechanism to compensate for suppression in the multi-strange sector. Instead, suppression in this sector is more efficiently accommodated through temperature variations, leading to moderate positive correlations between multi-strange eigenvolumes and $T_{\mathrm{ch}}$ ($\rho\sim0.25$--$0.35$). These trends illustrate the hierarchical role of different hadronic sectors in constraining bulk thermodynamic and geometric parameters.

	\subsubsection{ Yield Data  Fitting With Correlation}
	
	We investigate the impact of correlated systematic uncertainties in experimental hadron yield measurements on the Bayesian calibration of thermal-model parameters. A proper treatment of such correlations is essential, as neglecting them implicitly assumes statistical independence among data points and can lead to underestimated uncertainties or biased inference. In experimental yield measurements, systematic uncertainties commonly induce correlations across hadron species and centrality classes due to shared tracking, particle-identification, and reconstruction procedures. Since complete experimental covariance information is rarely available, we construct a phenomenological covariance matrix that captures the dominant intraspecies and centrality correlations based on physically motivated assumptions (details in Appendix~\ref{Error_Handling}). Our goal is not to reproduce the exact experimental covariance—a highly non-trivial task—but rather to quantify the sensitivity of thermal-model inference to the correlated systematics. The full GP predictive covariance is added to the phenomenological experimental covariance to construct the total covariance matrix used in the likelihood.
	
    
	To assess this sensitivity, we perform a controlled set of Bayesian calibrations in which the strength of the phenomenological correlation matrix is systematically varied. We consider three correlated scenarios: a weakly correlated case (WC), defined by scaling the baseline correlation matrix by a factor of 0.5 (BC$\times0.5$); a baseline correlated case (BC), corresponding to the construction described in Sec.~\ref{Error_Handling} (BC$\times1.0$); and a strongly correlated case (SC), obtained by scaling the baseline matrix by a factor of 1.5 (BC$\times1.5$). These scenarios are compared to a no-correlation (NC) reference case discussed in the previous section.
 
    Across all uncertainty treatments, posterior mean values remain stable, while inferred uncertainties, marginal shapes, and joint correlations exhibit a distinct non-monotonic and parameter-dependent response. Relative to the NC reference, the WC scenario shows a pronounced contraction of the 68\% credible intervals: freeze-out temperatures $T_{\mathrm{ch}}$ and system radii $R_{\mathrm{ch}}$ narrow by approximately $25$–$35\%$ and $20$–$30\%$, respectively, while eigenvolume parameters tighten by up to $\sim25\%$. In the BC scenario, temperature and radius constraints remain largely unchanged, whereas double-strange meson eigenvolume parameters broaden moderately. This trend intensifies in the SC limit, where eigenvolume-related credible intervals expand by $\sim20$–$75\%$, while local thermal parameters remain comparatively stable. 
    The observed behavior is supported by Jensen--Shannon divergence (JSD) measures~\cite{Endres2003JSDmetric}, which remain small overall ($\mathrm{JSD} \lesssim 0.06$), indicating that correlated systematic uncertainties primarily modulate inferential precision rather than induce substantial changes in posterior shape. The JSD for local parameters peaks in the WC scenario, while global parameters exhibit their largest divergence in the SC limit. Consistently, the Spearman rank correlation coefficients ($\rho^{\mathrm{SRC}}$)~\cite{Spearman:1904} show their strongest reorganization in the WC case ($|\Delta\rho^{\mathrm{SRC}}| \sim 0.25$--$0.30$), dominated by temperature--temperature and temperature--radius parameter pairs. At the baseline correlation strength, these changes persist with reduced magnitude ($|\Delta\rho^{\mathrm{SRC}}| \sim 0.20$--$0.25$), whereas in the SC scenario the dominant changes arise from enhanced correlations between eigenvolume parameters and system-size--related parameters $R_{\mathrm{ch}}$.
    
  
	 The observed non-monotonic response of inferred uncertainties and correlation structures reflects a transition between distinct statistical inference regimes as the strength of correlated systematics is varied. In the weakly correlated (WC) scenario, we observe a simultaneous contraction of credible intervals and a rapid reorganization of local parameter degeneracies, a pattern consistent with an inference regime in which moderate off-diagonal covariance terms suppress incoherent, statistically independent fluctuations that would otherwise be absorbed by a purely diagonal likelihood. In this regime, correlated uncertainties reduce the tendency of the model to fit uncorrelated noise across species and centrality bins, allowing shared yield information to be exploited more efficiently; consequently, parameter constraints sharpen and spurious statistical overfitting is mitigated, while posterior mean values remain largely unchanged.
	
	 As the correlation strength increases from the baseline (BC) toward the strong-correlation (SC) limit, the inference progressively enters a covariance-dominated regime. In the BC scenario, the near invariance of temperature and radius constraints indicates that local information gain has largely saturated, while the emerging broadening of eigenvolume parameters signals the onset of covariance control. This behavior intensifies in the SC limit, where strongly correlated systematic uncertainties substantially reduce the effective number of independent constraints, producing a marked expansion of credible intervals. In this regime, the joint posterior structure becomes increasingly governed by the imposed systematic covariance matrix $\boldsymbol{\Sigma}$ rather than by intrinsic parameter–parameter compensations, reflecting a redistribution of shared variance from parameter space into the covariance structure.	 
	
	 Additional insight is provided by the evolution of the reduced $\chi^{2}/\mathrm{ndf}$.  In the NC limit, the strongly sub-unity value ($\chi^{2}/\mathrm{ndf}\approx0.59$) is indicative of potential overfitting of statistical noise. Introducing weak correlations raises $\chi^{2}/\mathrm{ndf}$ to $\approx 0.97$, indicating a statistically well-calibrated likelihood. In the BC and SC scenarios, $\chi^{2}/\mathrm{ndf}$ decreases again, not due to improved physical agreement, but because enlarged correlated uncertainties relax the effective constraints and increase parametric freedom.
	 
	
	 Despite substantial changes in uncertainties and correlation structure, all qualitative physical conclusions remain robust. In particular, the inferred flavor-dependent hierarchy of eigenvolume interactions remains unchanged across all uncertainty treatments, demonstrating that it is not an artifact of covariance modeling but a robust feature within the flavor-based EV–HRG framework at chemical freeze-out. A further notable outcome is that even weakly correlated uncertainties act as an effective statistical noise filter, inducing substantial reorganization of marginal widths, posterior shapes, and parameter--parameter correlations, thereby underscoring the importance of explicitly incorporating experimental correlations. At the same time, the response to correlated systematics is strongly parameter dependent: local parameters saturate rapidly once correlations are introduced, whereas global eigenvolume parameters become increasingly sensitive as the correlation strength grows. Consequently, reliable inference requires not only the inclusion of experimental correlations but also a careful and systematic treatment of their structure in future studies. 

	\subsubsection{Simultaneous Fit to LQCD and Yield Data}
     
	Finally, we perform a unified Bayesian calibration of the Cross EV--HRG model by simultaneously incorporating LQCD thermodynamic observables and experimental hadron yield data. The primary objective of this joint analysis is to extract a common parameter set capable of consistently describing both datasets, thereby testing the consistency of the thermal description across fundamentally different regimes—the lattice-regulated grand canonical ensemble and the experimental freeze-out environment. In this calibration, the yield data incorporate the full experimental covariance matrix with baseline correlations (BC), while the LQCD sector utilizes a diagonal covariance matrix. The GP predictive covariance is included in both likelihoods to account for emulator uncertainty.
	
	Relative to the LQCD-only calibration, the combined LQCD+yield analysis exhibits a substantially enhanced constraining power, as anticipated from the yield-only results, while posterior mean shifts remain small for all parameters, indicating compatibility between the two datasets. The non-strange meson radius $r_M$ remains comparatively weakly constrained; however, its posterior width is significantly reduced relative to both the LQCD-only and yield-only calibrations, reflecting a statistical regularization of the meson sector in the simultaneous fit. The remaining eigenvolume parameters display a pronounced, sector-dependent response: for baryon and antibaryon excluded-volume parameters, the inclusion of yield data leads to substantial posterior reshaping, characterized by credible-interval reductions of order $85$–$90\%$ and observed Jensen--Shannon divergences ($\mathrm{JSD} \simeq 0.6$) relative to the LQCD-only reference. However, these JSD values are comparable to those observed in the yield-only calibration, indicating that the large divergence from the LQCD-only calibration reflects the strong information content of experimental hadron yield data rather than any physical tension between the two datasets. Consequently, yield data remain the primary source of constraint for baryonic eigenvolume parameters even in the presence of LQCD thermodynamic inputs, consistent with the strong sensitivity of abundant hadron yields to the baryonic repulsion scale.
 	 
	Focusing on the thermodynamic and geometric parameters, the transition from the yield-only fit to the combined LQCD+yield calibration produces only moderate, parameter-dependent effects. The freeze-out temperature parameters exhibit negligible and systematically positive shifts across all centrality classes, while the geometric radii show shifts of moderate magnitude that are predominantly negative. In both cases, the associated reductions in uncertainty remain modest and are accompanied by small Jensen--Shannon divergences. These results indicate that the inclusion of LQCD constraints leads to an incremental refinement of the freeze-out surface rather than a qualitative reshaping of the posterior distributions.
	
\begin{figure}
	\includegraphics[scale=0.4]{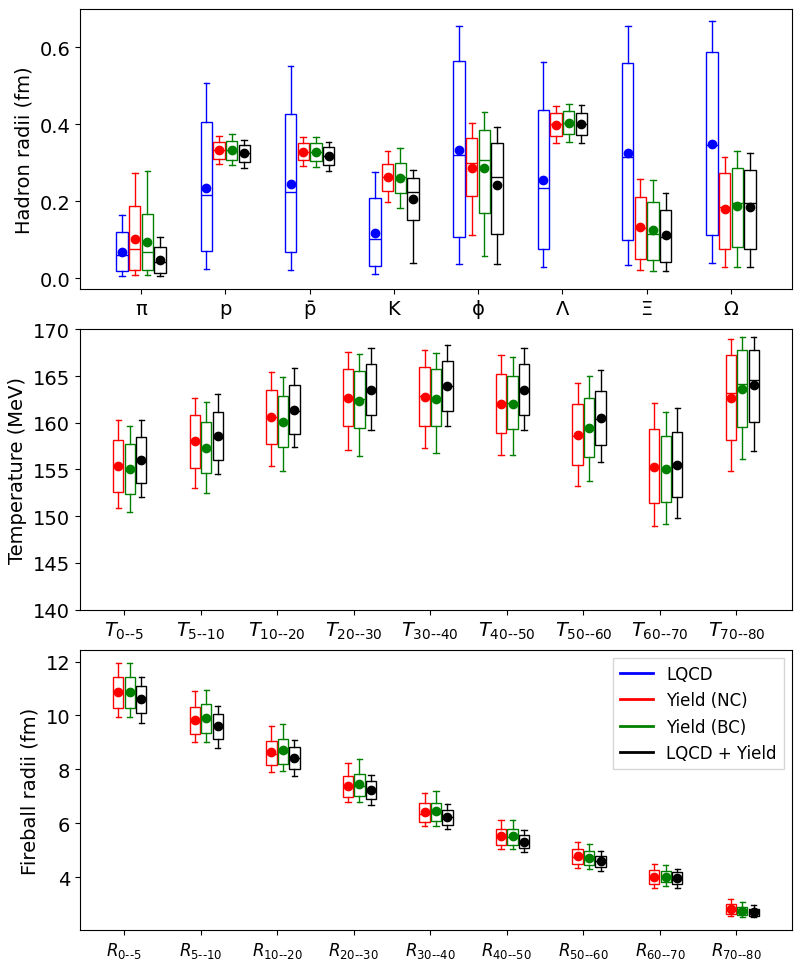}
	\caption{
		Comparison of model parameters extracted from different calibration setups: LQCD-only; yield-only without correlated systematic uncertainties (NC); yield-only with baseline correlated systematic uncertainties (BC); and the combined LQCD+yield calibration. The top panel shows the eigenvolume (hard-core) radii for the various hadron species. The middle and bottom panels display the centrality-dependent chemical freeze-out temperatures ($T$) and fireball radii ($R$), respectively.
		The datasets used in each calibration are indicated in the legend. Boxes denote the 68\% credible intervals, whiskers indicate the 90\% credible intervals, and circles mark the posterior mean values.
	}
	\label{fig:comparison_model_parameters}
\end{figure}

	
    Overall, the combined calibration preserves the qualitative physical interpretation of the model parameters, as illustrated by the comparison shown in Fig.~\ref{fig:comparison_model_parameters}, and provides clear support for the consistency of a single thermal description accommodating both LQCD thermodynamic observables and experimental hadron yield data within a common parameterization.

	
\section{Summary and Conclusions}
	\label{summary_conclusions}
     For the first time, LQCD thermodynamics and centrality-resolved hadron yield data from ALICE Pb--Pb collisions at $\sqrt{s_{\mathrm{NN}}}=2.76~\mathrm{TeV}$ are calibrated simultaneously using the Cross EV--HRG model within a unified Bayesian framework. The model incorporates flavor-dependent short-range repulsive interactions within a thermodynamically consistent formulation, with the corresponding eigenvolume radii treated as free parameters. The relative constraining power of the different datasets is systematically assessed. The effect of correlated experimental systematic uncertainties on parameter inference is rigorously studied by constructing a phenomenological covariance matrix and systematically varying its strength.
     
     The constraining power of the calibration is strongly dataset dependent. LQCD-only calibrations yield broad posterior distributions and weak constraints on most eigenvolume parameters, whereas hadron yield calibrations produce relatively sharply peaked posteriors and significantly tighter credible intervals on both flavor-resolved interaction parameters and centrality-dependent freeze-out parameters.
	
  	 Across all calibration strategies, the multi-strange eigenvolume parameters remain only moderately constrained, whereas the non-strange meson eigenvolume is weakly constrained and consistent with a vanishing effective radius within current uncertainties. The extracted baryon and antibaryon eigenvolumes are statistically consistent, as expected at vanishing baryon chemical potential ($\mu_B \simeq 0$).
  	 
 	 A clear flavor dependence of the eigenvolume parameters is indicated. The ordering inferred from the LQCD-only calibration differs from that obtained from hadron yield data; however, the LQCD-only results are strongly affected by large uncertainties. In contrast, the yield-driven hierarchy—although still subject to moderate uncertainties—remains stable across all yield-based calibration strategies, including the joint LQCD+yield analysis. Notably, in the strange-baryon sector this hierarchy does not follow the commonly adopted inverse-mass ordering assumed in earlier studies, motivating further investigations incorporating quark-model–predicted hadronic states and fluctuation-related observables. A comparison of the model parameters obtained from the different calibration strategies is presented in Fig.~\ref{fig:comparison_model_parameters}.
 	 
      The yield-based calibration effectively exposes intrinsic parameter degeneracies inherent to the thermal-model framework. Correlation analyses further demonstrate that all qualitative physical conclusions remain stable under different treatments of experimental systematic uncertainties, underscoring the robustness of the thermal interpretation within the flavor-dependent EV–HRG framework. At the same time, we find that the inclusion of correlated systematic uncertainties is important for avoiding artificial overfitting and for obtaining statistically meaningful uncertainty estimates. However, the impact of these correlations is strongly parameter- and strength-dependent, highlighting the need for a careful and systematic treatment of experimental correlations in precision thermal-model studies and Bayesian analyses of heavy-ion collision data.

	
      In conclusion, we establish a statistically robust Bayesian framework for thermal-model calibration and demonstrate that a unified thermal description of LQCD thermodynamics and experimental hadron yield data can be achieved within the Cross EV--HRG model. Several eigenvolume parameters, particularly in the strange sector, remain only weakly to moderately constrained. While no robust evidence for a strict inverse mass hierarchy of effective radii is observed, the present uncertainties do not allow for a definitive exclusion of such a hierarchy. This highlights the need for further dedicated studies to clarify the underlying structure of short-range repulsive interactions among hadrons. The analysis further shows that the inclusion of correlated experimental systematic uncertainties plays an important role in mitigating artificial overfitting and in obtaining statistically reliable uncertainty estimates. Finally, the presented framework provides a solid foundation that can be naturally extended to finite baryon chemical potential and to additional observables in future investigations

	\section{ACKNOWLEDGEMENTS}

	I am grateful to Partha Pratim Bhaduri, Ashik Iqbal, and V. Sreekanth for their meticulous review of the manuscript and insightful suggestions, and I would also like to thank Sandeep Chatterjee and the High Energy Physics group of IISER Berhampur for their generous support.

	\appendix

\section{Phenomenological Modeling of Systematic Uncertainties for Bayesian Calibration}
	\label{Error_Handling}
	
	
	\begin{figure}[H]
		\includegraphics[scale=0.4]{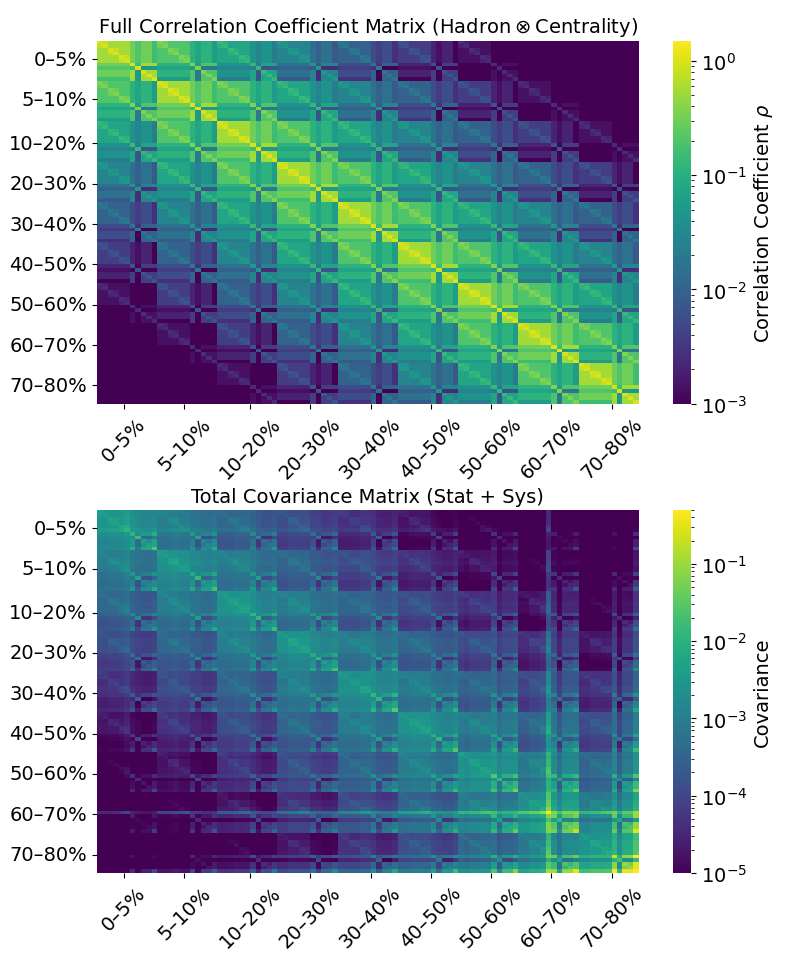}
		\caption{Phenomenologically constructed experimental uncertainty matrices for baseline case (BC). Top: systematic correlation matrix incorporating both centrality-bin and hadron species correlations. Bottom: total experimental covariance matrix, including both systematic and statistical uncertainties. Each block corresponds to a different hadron species.}
		\label{fig:exp_correlatation}
	\end{figure}
	
	\begin{table*} 
		\centering
		\caption{Qualitative correlation strengths among hadron yields due to shared sources of systematic uncertainties in Pb--Pb collisions at $\sqrt{s_{\mathrm{NN}}} = 2.76$. Categorization is based on shared reconstruction techniques, decay topologies, and detector responses. These values do not represent experimental extractions.}
		\begin{tabular}{lp{3cm}p{2.5cm}p{3.0cm}p{5.2cm}}
			\hline\hline
			\textbf{Category} & \textbf{Example Pairs} & \textbf{Correlation Type} & \textbf{Correlation Strength ($\rho$)} & \textbf{Systematic Source} \\
			\hline
			\hline
			\textbf{Particle–antiparticle} & $\pi^{+}$–$\pi^{-}$, $K^{+}$–$K^{-}$, $p$–$\bar{p}$, $\Lambda$–$\bar{\Lambda}$ & \textbf{Very Strong} & 0.60 -- 0.70 & Identical tracking, PID, acceptance, and selection strategies for charge-conjugate partners. \\
			
			\textbf{Primary light hadrons} & $\pi$–$K$, $\pi$–$p$, $K$–$p$ & \textbf{Strong} & 0.40 -- 0.50 & Common TPC tracking, TOF PID, similar vertexing; minor variation due to mass and resolution. \\
			
			\textbf{Light–strange} & $\pi$/$K$/$p$–$\Lambda$, $\pi$–$\Xi$, $p$–$\Xi$ & \textbf{Moderate} & 0.20 -- 0.30 & Partial overlap in tracking, strange hadrons use displaced vertex reconstruction. \\
			
			\textbf{Light– $\phi$--resonances} & $\pi$–$\phi$, $K$–$\phi$, $p$–$\phi$ & \textbf{Weak} & 0.10 -- 0.20 & Limited PID overlap.  \\
			
			\textbf{Strange–strange} & $\Lambda$–$\Xi$, $\Xi$–$\Omega$, $\Xi$–$\Lambda$ & \textbf{Strong} & 0.45 -- 0.50 & Common cascade/V0 reconstruction, displaced vertex PID, and topological selections. \\
			
			\textbf{Light–multi-strange} & $\pi$–$\Omega$, $p$–$\Omega$ & \textbf{Weak} & 0.10 -- 0.20 & Distinct decay topologies, minimal overlap in reconstruction or PID. \\

			\hline
		\end{tabular}
		\label{tab:correlation_matrix_strengths}
	\end{table*}
	
	Systematic uncertainties across different hadron species in heavy-ion collisions are correlated because many measurements rely on common reconstruction steps and detector responses. In the absence of published experimental correlation matrices, we construct a phenomenological covariance matrix informed by the known reconstruction characteristics of each species. The dominant sources of cross-species correlations include tracking efficiency, PID calibration, acceptance corrections, and material-budget or absorption effects. This construction is guided qualitatively by (i) ALICE detector-performance documentation~\cite{Abelev:2014ffa}, which details tracking, PID resolution, and acceptance corrections, and (ii) partial systematic uncertainty breakdowns from Pb--Pb hadron yield publications~\cite{Abelev:2013vea} and internal reports~\cite{ALICE-PUBLIC-2018-013}. Our aim is not to reconstruct the exact experimental correlations, but to assess how reasonable correlation assumptions affect the extracted thermal-model parameters. The qualitative correlation strengths adopted here, summarized in Table~\ref{tab:correlation_matrix_strengths}, should therefore be interpreted as phenomenological estimates.
 	
	Systematic correlations across different centrality bins are also expected. A widely used approach to model these correlations is to assume that the correlation coefficient \( \rho^{\text{cent}}_{ij} \) decays exponentially with the separation between bins~\cite{Soltz:2024gkm,JETSCAPE:2021ehl,Bernhard:2018hnz}:
	\begin{equation}
		\rho^{\text{cent}}_{ij} = \exp\!\left(-\frac{|i-j|}{\lambda}\right),
	\end{equation}
	where \( \lambda \) sets the centrality correlation length. In this work, we adopt \( \lambda = 1 \), following the choice in Ref.~\cite{Bernhard:2019bmu}.
	
 	
	In the absence of full experimental information, we adopt the simplified assumption that species-wise and centrality-wise systematic correlations factorize. We note, however, that some systematic effects may couple these dependencies, so this separability ansatz cannot capture all correlations exactly. More flexible covariance constructions may be explored in future work as additional experimental information becomes available.
	
	Under the separability assumption, the full correlation matrix is modeled as a Kronecker product:
	\begin{equation}
		\boldsymbol{\rho}^{\text{full}} 
		= \boldsymbol{\rho}^{\text{cent}} \otimes \boldsymbol{\rho}^{\text{had}},
	\end{equation}
	where \( \boldsymbol{\rho}^{\text{cent}} \) is the \(N_c \times N_c\) centrality correlation matrix and 
	\( \boldsymbol{\rho}^{\text{had}} \) is the \(N_h \times N_h\) hadron species correlation matrix. Here, 
	\( N_c \) denotes the number of centrality bins and \( N_h \) the number of hadron species.
	
	The full systematic covariance matrix is then constructed as
	\begin{equation}
		\boldsymbol{\Sigma}^{\text{sys}}
		= \mathbf{D}^{\text{sys}} \,
		\boldsymbol{\rho}^{\text{full}} \,
		\mathbf{D}^{\text{sys}},
	\end{equation}
	where \( \mathbf{D}^{\text{sys}} \) is a diagonal matrix whose entries are the
	systematic uncertainties associated with each yield measurement. In
	component form, this construction gives
	\begin{equation}
		\Sigma^{\text{sys}}_{(i,k),(j,l)}
		= \rho^{\text{cent}}_{ij} \,
		\rho^{\text{had}}_{kl} \,
		\sigma_{ik} \,
		\sigma_{jl},
	\end{equation}
	where \( \sigma_{ik} \) denotes the systematic uncertainty for hadron species
	\( k \) measured in centrality bin \( i \).
	
	Therefore, the total covariance matrix, considering both statistical and systematic uncertainties, is given by:
	\begin{equation}
		\boldsymbol{\Sigma}_{\text{exp}} = \boldsymbol{\Sigma}^{\text{sys}} + \boldsymbol{\Sigma}^{\text{stat}},
	\end{equation}
	where \( \boldsymbol{\Sigma}^{\text{stat}} \) is a diagonal matrix containing the statistical uncertainties for each measured hadron yield.
 	
	
	In the published ALICE yield data at $\sqrt{s_{\mathrm{NN}}}=2.76$~TeV, statistical and systematic components are not always reported separately. For these high-statistics Pb--Pb measurements, the statistical uncertainties are generally much smaller than the systematic ones. We therefore treat the quoted total uncertainties as being dominated by systematic effects, while still retaining the statistical contribution explicitly in the diagonal matrix $\boldsymbol{\Sigma}^{\text{stat}}$.
	
	Thus, the phenomenologically constructed correlation (top) and covariance (bottom) matrices are shown in Fig.~\ref{fig:exp_correlatation}. The total experimental covariance $\boldsymbol{\Sigma}_{\mathrm{exp}}$ (bottom panel) is then combined with the GP predictive covariance $\boldsymbol{\Sigma}_{\mathrm{GP}}$ and used in the MCMC likelihood (see Eq.~\ref{Eq:likelihood}).

\FloatBarrier

\end{document}